\documentclass[twocolumn,showpacs,showkeys,prb,floatfix,superscriptaddress]{revtex4}
\usepackage{graphicx}
\usepackage{bm}
\usepackage{amsmath}
\usepackage{amssymb}
\usepackage{array}

\begin{document}
\title{Optical orientation of electron spins in GaAs quantum wells}

\author{S. Pfalz}
\email{pfalz@nano.uni-hannover.de} \affiliation{Institut f\"{u}r
Festk\"{o}rperphysik, Universit\"{a}t Hannover, Appelstra{\ss}e 2,
D-30167 Hannover, Germany}

\author{R. Winkler}
\affiliation{Institut f\"{u}r Festk\"{o}rperphysik,
Universit\"{a}t Hannover, Appelstra{\ss}e 2, D-30167 Hannover,
Germany}

\author{T. Nowitzki}
\affiliation{Institut f\"{u}r Festk\"{o}rperphysik,
Universit\"{a}t Hannover, Appelstra{\ss}e 2, D-30167 Hannover,
Germany}

\author{D. Reuter}
\affiliation{Lehrstuhl f\"{u}r Angewandte Festk\"{o}rperphysik,
Ruhr-Universit\"{a}t Bochum, Universit\"{a}tsstra{\ss}e 150,
D-44780 Bochum, Germany}

\author{A. D. Wieck}
\affiliation{Lehrstuhl f\"{u}r Angewandte Festk\"{o}rperphysik,
Ruhr-Universit\"{a}t Bochum, Universit\"{a}tsstra{\ss}e 150,
D-44780 Bochum, Germany}

\author{D. H\"{a}gele}
\affiliation{Institut f\"{u}r Festk\"{o}rperphysik,
Universit\"{a}t Hannover, Appelstra{\ss}e 2, D-30167 Hannover,
Germany}

\author{M. Oestreich}
\affiliation{Institut f\"{u}r Festk\"{o}rperphysik,
Universit\"{a}t Hannover, Appelstra{\ss}e 2, D-30167 Hannover,
Germany}

\date{\today}

\begin{abstract}
  We present a detailed experimental and theoretical analysis of the
  optical orientation of electron spins in GaAs/AlAs quantum wells.
  Using time and polarization resolved photoluminescence excitation
  spectroscopy, the initial degree of electron spin polarization is
  measured as a function of excitation energy for a sequence of
  quantum wells with well widths between $63$~{\AA} and $198$~{\AA}.
  The experimental results are compared with an accurate theory of
  excitonic absorption taking fully into account electron-hole
  Coulomb correlations and heavy-hole light-hole coupling. We find
  in wide quantum wells that the measured initial degree of
  polarization of the luminescence follows closely the spin
  polarization of the optically excited electrons calculated as a
  function of energy. This implies that the orientation of the
  electron spins is essentially preserved when the electrons relax
  from the optically excited high-energy states to quasi-thermal
  equilibrium of their momenta. Due to initial spin relaxation, the
  measured polarization in narrow quantum wells is reduced by a
  constant factor that does not depend on the excitation energy.
\end{abstract}

 \maketitle
\section{\label{sec:intro}Introduction}

The optical excitation of semiconductors with circularly polarized
light creates spin-polarized electrons in the conduction band.
\cite{dyakonov84} The degree of electron spin polarization
obtainable by means of \emph{optical orientation} can reach almost
$100$\%, depending on the conduction and valence band states
involved in the optical transition. The intimate relation between
electron spin and circularly polarized light has formed the basis
for many of the pioneering experiments of semiconductor
spintronics. Optical investigations demonstrated the efficient
injection of spin polarized
electrons,\cite{Oestreich99,Fiederling99} the transport of spin
polarized electrons over macroscopical
distances,\cite{Haegele98,Kikkawa99} manipulation and storage of
spin orientation,\cite{salisNATURE2001} and the interaction with
nuclear momenta \cite{lampelPRL68}. Furthermore, the spin
dependence of optical transitions can be utilized to switch the
intensity and polarization of a semiconductor laser by changing
the spin orientation of injected electrons.\cite{hallsteinPRB97}
Recently, the reduction of the threshold in semiconductor lasers
pumped with spin-polarized electrons was observed.\cite{Rudolph03}
But although optical orientation has proven to be a powerful tool
to study electron spins in quasi-two-dimensional (quasi-2D)
semiconductor systems, the present understanding of spin
orientation is based on crude approximations. A direct comparison
of experimentally determined degrees of spin orientation with an
accurate theoretical treatment is still missing. The goal of this
paper is thus to present a systematic experimental and theoretical
study of the optical orientation of electron spins in quasi-2D
systems.

\begin{figure}[t]
\includegraphics[width=0.9\linewidth]{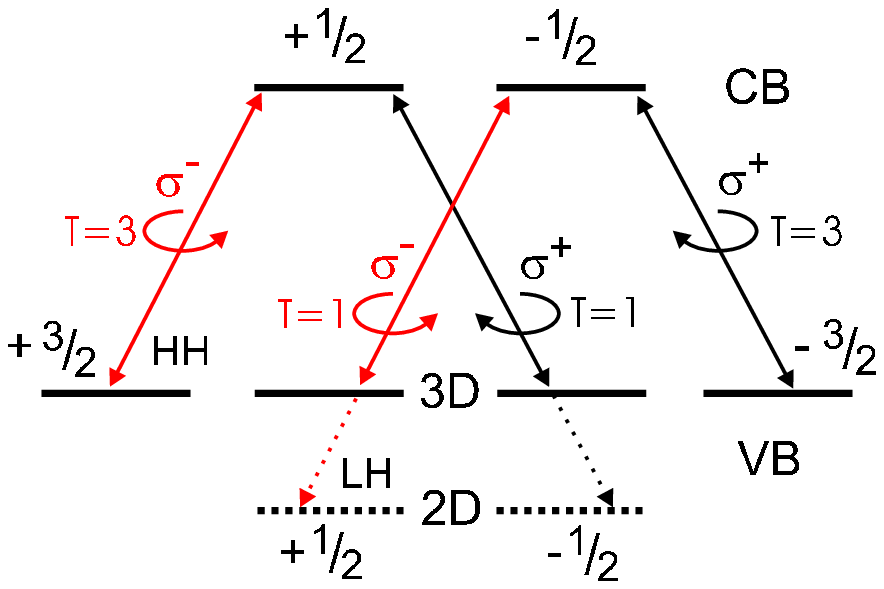}
\caption{\label{fig:selectrule} Selection rules and
relative transition rates $T$ for optical transitions between
valence band (VB) states having an effective spin $S=3/2$ and
conduction band (CB) states with $S=1/2$
(Ref.~\onlinecite{dyakonov84}). In bulk semiconductors the HH
states ($S_z = \pm 3/2$) are degenerate with the LH states ($S_z =
\pm 1/2$) whereas in quasi-2D systems the LH states (dotted bold
lines) are lower in energy than the HH states.}
\end{figure}

In direct semiconductors like GaAs, the selection rules for optical
transitions from the uppermost valence band to the lowest conduction
band are commonly based on the simple picture that the electron
states in the conduction band have spin $S=1/2$ whereas the hole
states in the valence band have an effective spin $S=3/2$. The hole
states with spin $z$-component $S_z = \pm 3/2$ are denoted
heavy-hole (HH) states whereas the light-hole (LH) states have $S_z
= \pm 1/2$. For absorption and emission of circularly polarized
light we thus get the selection rules depicted in
Fig.~\ref{fig:selectrule} (Ref.~\onlinecite{dyakonov84}). According
to this scheme, the transition probability from the HH states to the
conduction band is three times larger than from the LH states. In
bulk semiconductors, we thus expect that the maximum attainable
degree of spin polarization is $P_s=0.5$, where $P_s$ is defined as
\begin{equation}\label{eq:PolgradAllgemein}
P_s = \frac{N_+ - N_-}{N_+ + N_+}
\, ,
\end{equation}
and $N_+$ ($N_-$) is the number of electrons with spin up (down),
respectively. In 2D systems the degeneracy of the HH and LH states
is lifted as sketched in Fig.~\ref{fig:selectrule}. For resonant
excitation at the HH energy we thus expect a rise of the maximum
attainable degree of polarization up to $P_s = 1$.

Even in a single-particle picture for the optical
excitation, the naive $3:1$ ratio of HH and LH transitions is
obtained only if HH-LH coupling of the hole states at nonzero wave
vectors $\bm{k}$ is neglected. Due to this HH-LH coupling, the hole
states with $k>0$ are not spin eigenstates. Furthermore, a realistic
treatment must take into account that optical absorption gives rise
to the formation of excitons, i.e., Coulomb correlated
electron-holes pairs. Thus even for excitations close to the
absorption edge we get substantial HH-LH coupling because the
exciton states consist of electron and hole states with $k$ of the
order of $1/a_\mathrm{B}^\ast$, where $a_\mathrm{B}^\ast$ is the
effective Bohr radius. The Coulomb coupling between electron and
hole states yields a second contribution to the mixing of single
particle states with different values of $S_z$. Finally, we must
keep in mind that for higher excitation energies we get a
superposition of exciton continua that are predominantly HH- or
LH-like. These different excitons contribute oppositely to the spin
orientation of electrons. We note that these arguments are valid for
the optical excitation of bulk semiconductors and quasi-2D systems.

In early works, several groups \cite{Weisbuch81, Masselink84}
reported on polarization resolved transmission and
photoluminescence (PL) experiments on GaAs/AlGaAs quantum wells
(QWs) under cw excitation. They measured the polarization as a
function of excitation energy for a small range of excess
energies. In later works, the electron spin polarization $P_s$ in
quasi-2D systems was studied using time-resolved photoluminescence
excitation spectroscopy. For excitation energies even slightly
above the HH resonance, several authors \cite{Freeman90, Dareys93,
Munoz95} observed a polarization $P_s$ that was significantly
smaller than one. These measurements were carried out on fairly
narrow GaAs/AlGaAs multiple QWs with well widths $w=40$~{\AA}
(Refs.\ \onlinecite{Freeman90, Dareys93}) and $77$~{\AA} (Ref.\
\onlinecite{Munoz95}). A first well-width dependent study of
optical orientation was performed experimentally by Roussignol
{\it et al.}, \cite{Roussignol92} but only for excitation energies
up to $30$~meV above the HH resonance. For energies near the HH
resonance, Roussignol {\it et al.} found initial spin
polarizations in the range $60-80$\%, whereas they expected values
between $85$ and $90$\%. They argued that additional relaxation
mechanisms were required to describe their results. Kohl {\it et
al.} \cite{Kohl91} studied the optical orientation in an
$80$~{\AA} wide GaAs QW for an excess energy of $\Delta E =
60$~meV above the HH absorption edge. In contrast to our findings
discussed below, they observed for this value of $\Delta E$ a
rather large initial spin polarization close to $100$\%.

Twardowski and Hermann \cite{Twardowski87} as well as Uenoyama and
Sham \cite{Uenoyama90} studied the polarization of QW PL
theoretically, taking into account HH-LH coupling in the valence
band. However, these authors neglected the Coulomb interaction
between electron and hole states. On the other hand, Maialle {\em
et al.} \cite{Maialle93} investigated the spin dynamics of
excitons taking into account the exchange coupling between
electrons and holes, but they disregarded the HH-LH coupling in
the valence band. Both the HH-LH coupling and the Coulomb coupling
are known to be important for an accurate description of excitonic
spectra. \cite{Winkler95} To the best of our knowledge, no
systematic experimental or theoretical examination of $P_s$ for
different well widths and a wide range of excitation energies has
been reported so far.

In this work we experimentally analyze the energy dependence of
the optical selection rules for the creation and recombination of
spin polarized carriers by investigating the time-dependent
polarized luminescence of seven $(100)$ GaAs/AlAs QWs with well
widths from $63$ to $198$~{\AA} and excitation energies between
$1.529$ and $1.744$~eV. We compare these results with an accurate
theory of excitonic absorption taking into account Coulomb
coupling and HH-LH coupling between the subbands. \cite{Winkler95}
The experimental results for a wide range of parameters are in
good agreement with the parameter-free calculations. We find that
the measured initial optical polarization of the luminescence
follows closely the spin polarization of the optically excited
electrons calculated as a function of energy. This implies that
the orientation of the electron spins is essentially preserved
when the electrons relax from the optically excited high-energy
states to quasi-thermal equilibrium of their momenta. In narrow
QWs, however, the measured polarization is reduced due to fast
initial spin relaxation that is almost independent of the
excitation energy.

The paper is organized as follows. Section~\ref{sec:setup} describes
the experimental setup and the sample under investigation. In
Sec.~\ref{sec:expresults}, we first present the results for a
$198$~{\AA} wide QW where we obtain very good agreement between
experiment and theory. Second, we discuss how the polarization
observed in narrow QWs is reduced because of fast initial spin
relaxation directly after laser excitation. Our theory for optical
orientation is introduced in Sec.~\ref{sec:theory}, where we give a
detailed discussion of the influence of Coulomb coupling and HH-LH
coupling for an accurate theoretical description of the optical
orientation of electron spins. The conclusions are summarized in
Sec.~\ref{sec:conclusion}.

\section{\label{sec:setup}Experimental Methods}

The sample under investigation is a high quality intrinsic
GaAs/AlAs structure containing twelve single QWs with different
well widths grown by MBE on a $(100)$ oriented GaAs substrate.
\cite{Sogawa01} The QWs are separated by a triple layer of
$26$~{\AA} AlAs, $27$~{\AA} GaAs, and $26$~{\AA} AlAs. In this
work we present experimental data for the seven broadest QWs with
well widths between $63$ and $198$~{\AA}.\cite{wellsNOTE} The
sample is mounted in a finger cryostat and all measurements were
performed at a temperature of $4.2$~K. Pulses from a Kerr-lens
mode-locked Ti:sapphire laser excite the sample with a repetition
rate of $80$~MHz. We use a pulse shaper to reduce the spectral
linewidth of the $100$~fs pulses to $0.8$~nm full width at half
maximum (FWHM). The wavelength is tuned from $711$~{nm} to
$811$~{nm} in steps of 1~nm. The maximum excitation power is
limited to about $2$~mW, because most of the laser power is
blocked by the pulse shaper. We estimate that the optically
excited carrier density lies in the range $2\times 10^{8} -
2\times 10^{9}$~cm$^{-2}$ depending on the QW width and the
excitation energy. We carefully control the polarization of the
exciting laser pulse by means of a Soleil-Babinet polarization
retarder, taking into account the dependence of the retardation on
the excitation wavelength. The retarder is readjusted for each
excitation wavelength to achieve close to 100\% circularly
polarized light. The PL is measured in reflection geometry by a
synchroscan streak camera providing a spectral and temporal
resolution of $7$~meV and $15$~ps, respectively. We separately
detect the two circularly polarized PL components $\sigma^\pm$
using an electrically tunable liquid-crystal retarder. Each QW
emits light only at its energetically lowest excitonic resonance.
Since the PL wavelengths of the QWs vary over a wide range and the
liquid-crystal retarder shows a chromatic dependence of the
retardation, the PL data were corrected independently for each QW
according to the measured dispersion curve of the retarder.

\begin{figure}[t]
\includegraphics[width=0.9\linewidth]{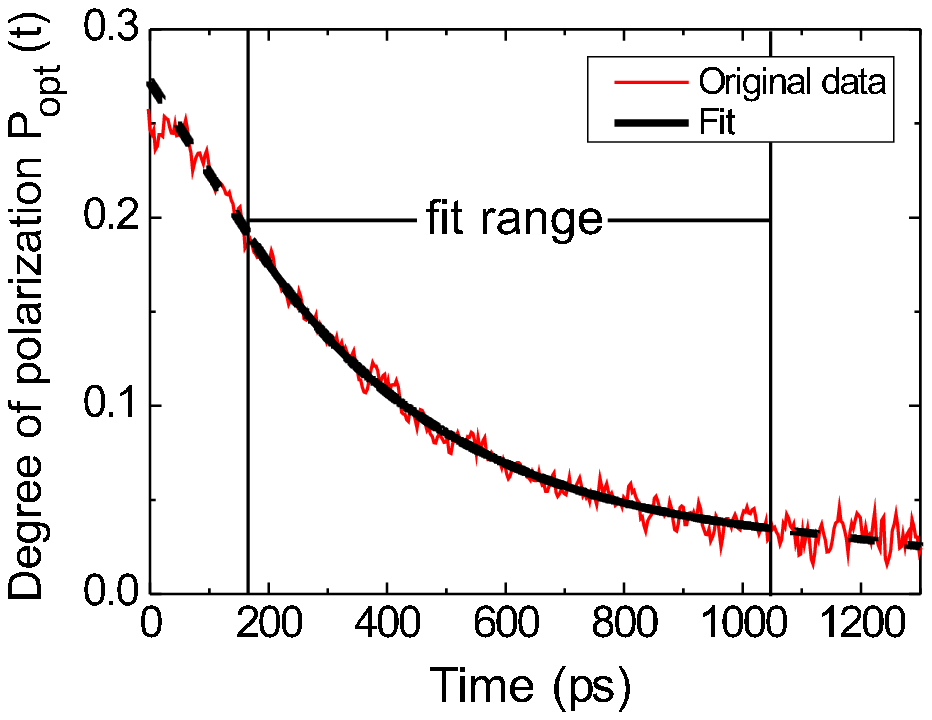}
\caption{\label{fig:DegPol} Time dependent degree of
polarization $P_\mathrm{opt}(t)$ for the $152$~{\AA} wide QW at an
excitation energy of $1.744$~eV (red line) and exponential fit
based on Eq.\ (\ref{eq:polfit}) (black line).}
\end{figure}

We obtain the time dependent degree of optical polarization
\begin{equation}
P_\mathrm{opt}(t)=\frac{I_+(t) - I_-(t)}{I_+(t) + I_-(t)}
\end{equation}
from the time resolved PL spectra, where $I_\pm (t)$ is the PL
intensity of the $\sigma^\pm$ component. $P_\mathrm{opt}(t)$ is
measured for each QW scanning the excitation energy from $1.529$
to $1.744$~eV. As an example, Fig.~\ref{fig:DegPol} shows
$P_\mathrm{opt}(t)$ for the $152$~{\AA} QW at an excitation energy
of $1.744$~eV which corresponds to an excess energy of $209$~meV
above the lowest HH resonance. We determine the initial degree of
polarization by fitting $P_\mathrm{opt}(t)$ to
\begin{equation}
  \label{eq:polfit}
  P_\mathrm{opt}(t) = P_1 + P_0 \exp (-t/\tau_s) \, ,
\end{equation}
where $\tau_s$ is the decay time of $P_\mathrm{opt}(t)$. We
identify $P_0$ with the optical polarization at $t = 0$. $P_1$
corresponds to an offset in the measurement of usually below
$0.02$ which is probably due to a slight linear polarization
introduced by the liquid crystal retarder. The error $P_1$ is
included in the error bars of $P_0$.

The central idea underlying the interpretation of our experiments is
that we can identify the measured degree of optical polarization
with the electron spin polarization, $P_\mathrm{opt} (t) = P_s(t)$.
This association is based on the following arguments. First we
recall that the electron spin relaxation is usually slow compared to
the hole spin relaxation. \cite{Damen91} Therefore, every electron
can radiatively recombine with an appropriate hole state. Second we
note that the measured PL reflects only the HH1:E1(1s) transition.
[In this paper we label optical transitions by the hole (HH or LH)
and electron (E) subbands contributing dominantly to the excitonic
states. For a bound exciton we append in brackets the quantum number
of the bound state. \cite{Winkler95} See also the discussion in
Sec.~\ref{sec:theory}.] Our calculations indicate that for this
transition we have a strict one-to-one correspondence between the
spin polarization and the degree of optical orientation, with
completely spin polarized electrons giving rise to perfectly
circularly polarized light. This is confirmed by the experiments
showing a very high degree of optical polarization for the HH1:E1
transition. Third, we assume that the electron spin polarization is
preserved during the first few ps after laser excitation while the
electrons relax from the optically excited high-energy states to
quasi-thermal equilibrium for the momenta. This assumption is best
fulfilled in wide QWs, see our discussion of initial spin relaxation
in Sec.~\ref{subs:InitialRelaxation}. Finally, we remark that the
above arguments imply that the decay time $\tau_s$ in Eq.\
(\ref{eq:polfit}) can be identified with the spin relaxation time of
the electrons.

\section{\label{sec:expresults}Results and discussion}

\subsection{\label{subs:ExpTheo}Initial degree of polarization}

\begin{figure}[t]
\includegraphics[width=0.9\linewidth]{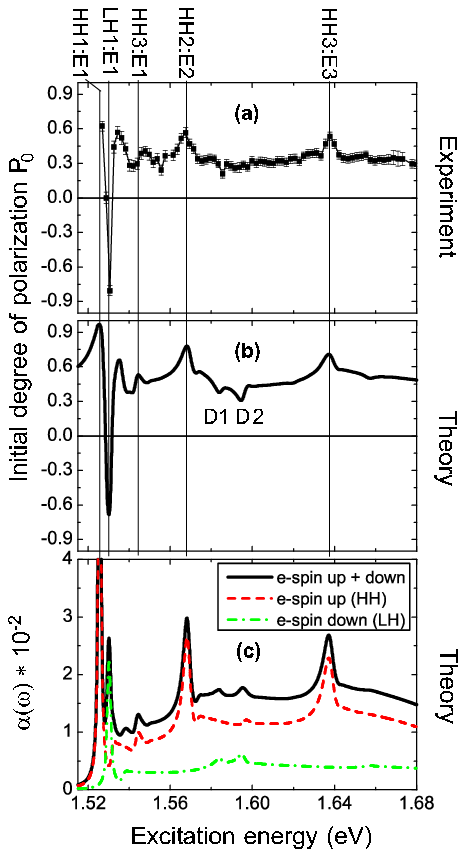}
\caption{\label{fig:198} (a) Measured and (b)
calculated initial degree of spin polarization $P_0$ as a function
of excitation energy of the $198$~{\AA} wide GaAs/AlAs QW. The
vertical lines label the resonances in (a) and (b) according to
the dominantly contributing electron (E) and hole (LH or HH)
subbands as discussed in Sec.~\ref{sec:theory}. The black line in
(c) shows the calculated absorption coefficient $\alpha(\omega)$.
The red dashed line (green dash-dotted line) in (c) shows how
excitons containing spin up (spin down) electrons contribute to
$\alpha(\omega)$.}
\end{figure}

In this section we will discuss optical spin orientation for the
$198$~{\AA} QW. Here, the excitation power is $200$~$\mu$W and the
laser spot radius is approximately $125~\mu$m which creates a low
carrier density of the order of $5\times 10^8$~cm$^{-2}$. We choose
this low excitation power to avoid a spectral overlap of the PL from
the substrate with the PL from the QW. Figure~\ref{fig:198} shows
the measured [Fig.~\ref{fig:198}(a)] and the calculated
[Fig.~\ref{fig:198}(b)] degree of spin polarization as a function of
excitation energy. For comparison, the solid line in
Fig.~\ref{fig:198}(c) shows the calculated absorption spectrum
$\alpha (\omega)$ for $\sigma^-$ polarized light. While the solid
line contains contributions from all dipole allowed exciton states
at energy $\hbar \omega$, the broken lines differentiate between the
contributions of those states to $\alpha (\omega)$, whose electron
spin is oriented either up or down. In agreement with
Fig.~\ref{fig:selectrule}, these contributions are essentially the
same as the contributions of HH and LH states to $\alpha(\omega)$.
The partitioning of $\alpha(\omega)$ combined with the calculated
electron and hole subband energies allows us to label the peaks in
the polarization spectra in Figs.~\ref{fig:198}(a) and
\ref{fig:198}(b) by the electron (E) and hole (HH or LH) subbands.
The solid vertical lines in Fig.~\ref{fig:198} indicate the
identified peaks. A more detailed discussion of the labeling scheme
will be given in Sec.~\ref{sec:theory}.

The first positive peak at 1.525~eV corresponds to the HH1:E1(1s)
transition. Next we find a narrow region around 1.53~eV with
negative $P_0$ which we attribute to the LH1:E1(1s) transition.
For the 198~{\AA} wide QW investigated here the LH1:E1(1s) exciton
is below the continuum of HH1:E1 excitons so that the LH1:E1(1s)
exciton is a discrete state (i.e., not a Fano resonance).
Therefore, $|P_0|$ is smaller than one only due to the homogenous
broadening of the exciton states. The next peak at 1.535~eV
reflects the absorption edge of the HH1:E1 exciton continuum. The
LH component in Fig.~\ref{fig:198}(c) exhibits a minimum which
explains the large positive value of $P_0$. The LH1:E1 absorption
edge at 1.539~eV, along with the decreasing HH contribution, leads
to a reduction of $P_0$ followed by a peak at 1.544~eV which
corresponds to the HH3:E1(1s) transition. We attribute the
following peak at 1.568~eV to the HH2:E2(1s) transition, while the
peak at 1.637~eV corresponds to the HH3:E3(1s) transition. All
these structures are found in both experiment and theory.

Interestingly, theory shows a dip of $P_0$ at 1.584~eV labeled D1
in Fig \ref{fig:198}(b)  which appears to be related to a
transition from an LH state to the conduction band. However,
unlike the transitions discussed above, it cannot be related to a
particular pair of electron and hole subbands. Furthermore, we
observe a dip of $P_0$ at 1.595~eV (D2) in Fig.~\ref{fig:198}(b).
The calculations indicate that two excitons are almost degenerate
at this energy, the LH2:E2(1s) and the HH4:E2 exciton with the
latter being slightly higher in energy. (We remark that, strictly
speaking, all excitons above the HH1:E1 absorption edge at
$1.535$~eV are Fano resonances. \cite{Winkler95})

Finally we note that there is a very good agreement between
experiment and theory not only for the individual features in the
spectra, but also for the relative height of the peaks and the
general trends of $P_0$ as a function of energy. The good
agreement between the experimental data and the calculated results
shows that our theory provides a realistic picture of the spin and
energy dependent optical selection rules in GaAs QWs. In
Sec.~\ref{sec:theory} we show that it is vital for our
quantitative theory to take into account both Coulomb coupling and
HH-LH coupling. If these couplings are neglected, we get
substantial deviations between experiment and theory.

\subsection{\label{subs:InitialRelaxation}Initial spin relaxation}

\begin{figure*}[p]
\includegraphics[width=0.88\linewidth]{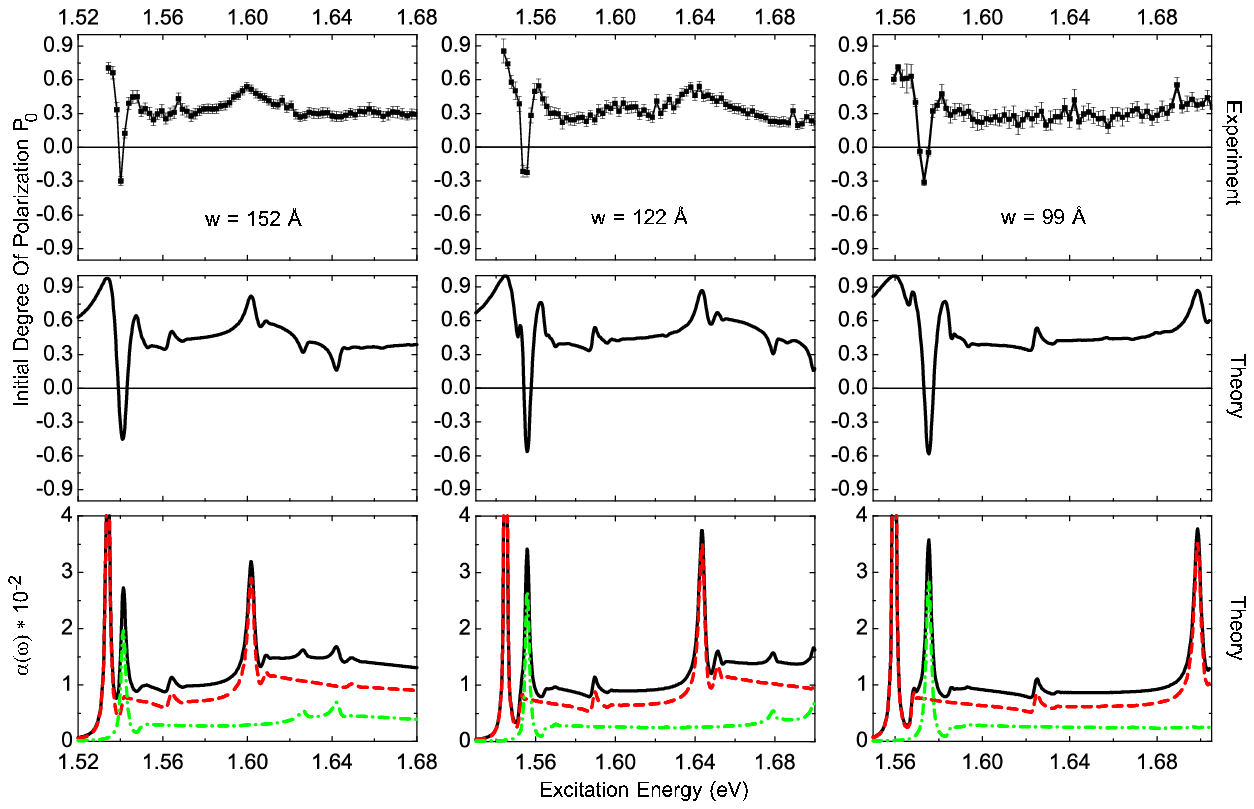}
\vspace{3mm}
\includegraphics[width=0.88\linewidth]{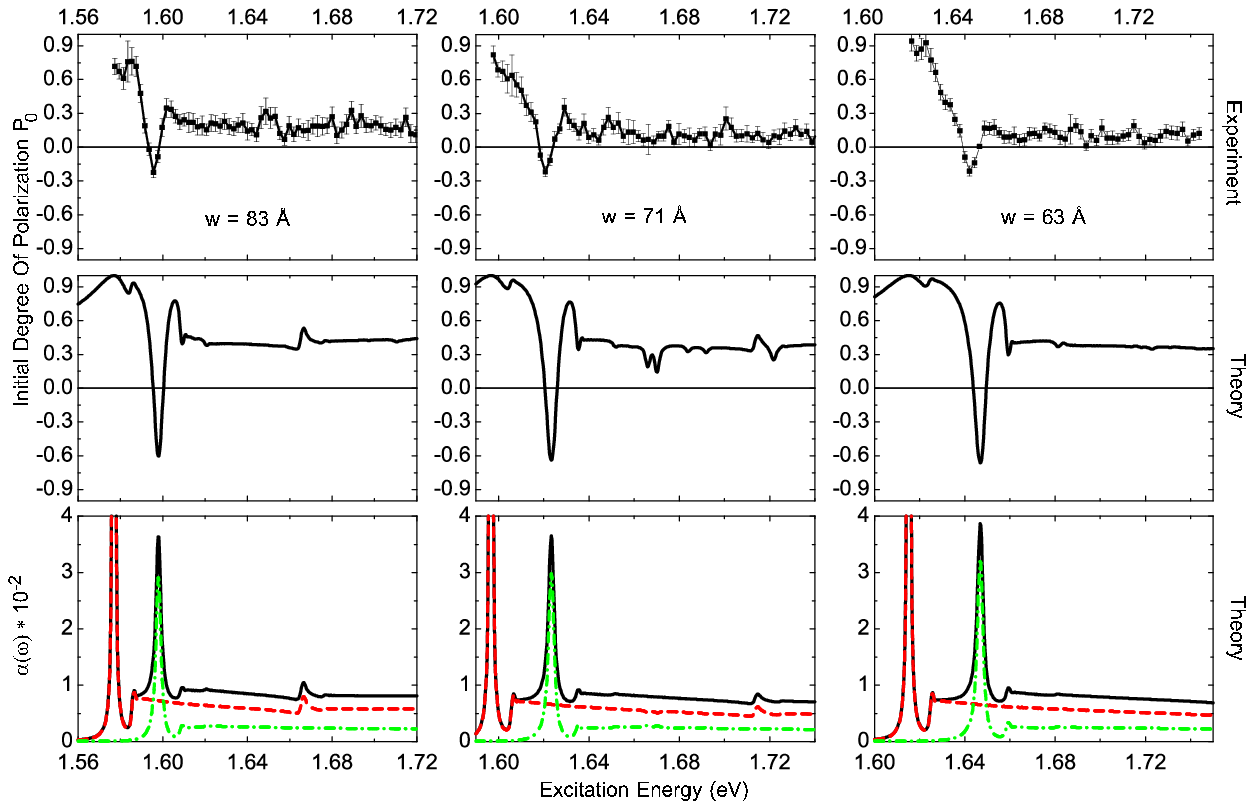}
\caption{\label{fig:AllQWs} Initial degree of spin
polarization $P_0$ as a function of excitation energy for
GaAs/AlAs QWs with different well widths. The upper panels show
experimental data measured at an excitation power of 1~mW, the
central panels are calculated results. For comparison, the black
lines in the lower panels show the calculated absorption
coefficient $\alpha(\omega)$. The red dashed line (green
dash-dotted line) in the lower panels show how excitons containing
spin up (spin down) electrons contribute to $\alpha(\omega)$, see
Eq.\ (\ref{oszi_spinor}).}
\end{figure*}

In Sec.~\ref{sec:setup} we identified the measured degree of optical
polarization with the spin orientation of excited electrons based on
the assumption that the electron spin polarization is preserved when
the electrons relax from high-energy states to thermal equilibrium.
This assumption is well fulfilled for wide QWs (see
Fig.~\ref{fig:198}) where we obtained good agreement between
absolute values of the measured and calculated spin polarization
$P_0$. Figure~\ref{fig:AllQWs} shows the measured and calculated
$P_0$ for the QWs with well widths between $152$ and $63$~{\AA} and
an excitation power of $1$~mW.
Once again, we obtain good agreement between experiment and theory
for many features in the spectra. The ratio between measured and
calculated values decreases, however, with decreasing well width.
The ratio is close to one for the $198$~{\AA} wide QW, but becomes
much smaller for the narrow QWs. Interestingly, this ratio is for
each QW approximately constant for a large range of energies. We
propose that the reduced value of $P_0$ is due to fast initial
spin relaxation of the excited electrons prior to establishing
thermal equilibrium for their momenta.

We assume that this mechanism is similar to the Dyakonov-Perel
(DP) spin relaxation of electrons. \cite{dyakonov72} An excitation
with energies above the HH1:E1 resonance creates electrons with
large wave vectors $\bm{k}_\|$. In these states, the electron spins
are exposed to an effective magnetic field $\bm{\Omega} (\bm{k}_\|)$
due to the conduction band spin splitting. While the electrons relax
from the excited states to states in thermal equilibrium with
smaller wave vectors $\bm{k}_\|$, the spins precess around the field
$\bm{\Omega} (\bm{k}_\|)$ so that the measured spin orientation is
reduced. We call this process initial spin relaxation.
\cite{Munoz95} We note that, in general, DP spin relaxation becomes
more efficient for larger electron energies so that the time scale
of the initial spin relaxation is shorter than the spin relaxation
time $\tau_s$ at later times (compare Figure~\ref{fig:DegPol}).

To obtain a qualitative estimate of how the initial spin
relaxation influences the measured polarization $P_0$, we evaluate
the average spin precession period $T_z$ of the optically excited
electron states prior to the first scattering event. In the
following, we consider only inelastic scattering processes with
energy relaxation time $\tau_E$ and neglect the motional narrowing
so that the calculated precession period $T_z$ is a lower bound
for the time scale of the inital spin relaxation (cf.\
Sec.~\ref{sec:Density}). Furthermore, we neglect Coulomb coupling
so that the electron states can be characterized by the in-plane
wave vector $\bm{k}_\|$.

We start with the expression for the effective magnetic field vector
in symmetric (100)-oriented QWs \cite{Kainz03}
\begin{equation}\label{FieldVector}
\bm{\Omega} (\bm{k}_\|) = \frac{2\gamma}{\hbar} \left(
  \begin{array}{c}
  k_x \left(k_y^2 - \langle k_z^2\rangle\right) \\[0.5ex]
  k_y \left(\langle k_z^2\rangle - k_x^2\right) \\[0.5ex]
  0
  \end{array} \right),
\end{equation}
where $\bm{k}_\| = (k_x, k_y, 0)$ is the in-plane wave vector, $k_z$
is the quantized perpendicular component of $\bm{k}$, and $\gamma$
is the Dresselhaus coefficient. For an isotropic dispersion we
obtain the average precession frequency $\langle \Omega_z \rangle
(k_\|)$ of electron spins polarized in $z$ direction by averaging
$|\bm{\Omega} (\bm{k}_\|)|$ over the polar angle~$\phi$ of
$\bm{k}_\| = k_\| (\cos \phi, \sin \phi, 0)$
\begin{equation}\label{omega_max}
\langle \Omega_z \rangle (k_\|) = \frac{1}{2\pi}\int_0^{2\pi}
\mathrm{d}\phi \, \left|\Omega\left(\bm{k}_\|\right)\right|.
\end{equation}
Assuming a parabolic dispersion of the electron excess energy,
$\Delta E = \hbar^2 k_{\|}^2 / (2 m^\ast)$ with effective mass
$m^\ast$, we can express the precession period $T_z = 2\pi / \langle
\Omega_z \rangle$ in terms of $\Delta E$.

The quantity $T_z$ provides an estimate for the timescale on which
the optically induced spin orientation is lost. It competes with the
timescale $\tau_E$ of the energy relaxation. We can estimate the
ratio between the measured and the optically excited spin
polarization by calculating
\begin{subequations}
\label{cosine}
\begin{eqnarray}
R (k_\|) & = &
\begin{array}[t]{>{\displaystyle}r}
\frac{1}{\tau_E (k_\|)}
\int \frac{\mathrm{d} \phi}{2\pi} \int_0^{\infty} \mathrm{d}t \,
\exp\left[-t / \tau_E (k_\|) \right]
\\ \times
\cos\left[|\Omega (\bm{k}_\|)| \, t\right]
\end{array} \\
& = & \int \frac{\mathrm{d} \phi}{2\pi} \;
\frac{1}{1 + |\Omega (\bm{k}_\|)|^2 \, \tau_E^2 (k_\|)} \,,
\end{eqnarray}
\end{subequations}
where we have assumed that the occupation of the initially excited
states $\bm{k}_\|$ decreases exponentially with decay time
$\tau_E$. For $\tau_E \ll T_z$ we obtain $R (k_\|) \approx 1 -
(2\pi\tau_E/T_z)^2$.

\begin{figure}[t]
\includegraphics[width=0.85\linewidth]{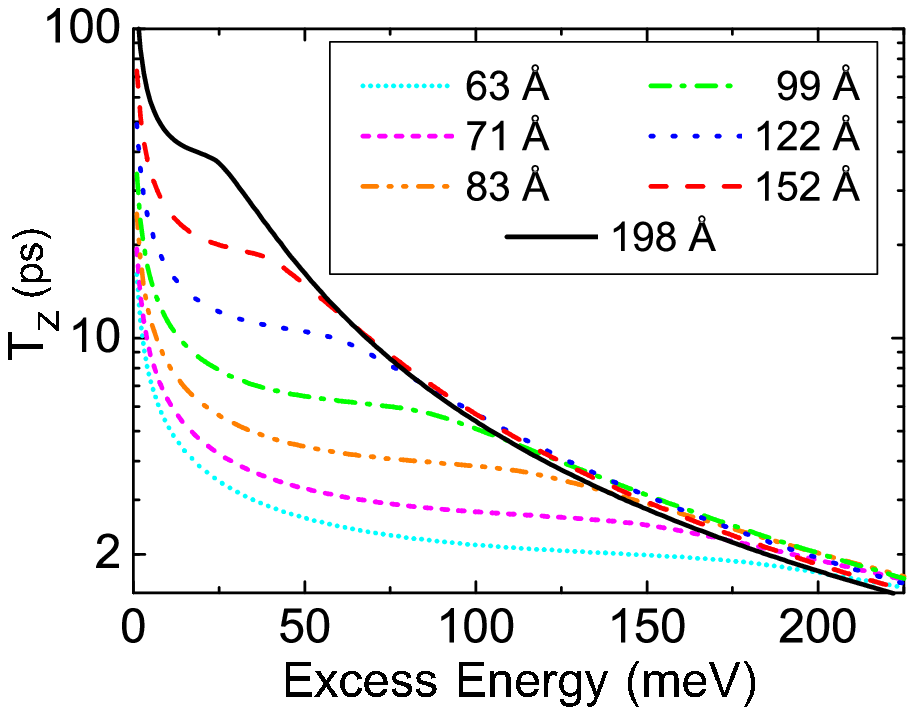}
\caption[]{\label{fig:Calculation} Average spin
precession time $T_z = 2\pi / \langle \Omega_z \rangle$ as a
function of excess energy $\Delta E$ for different QWs with well
widths between $63$ and $198$~{\AA}.}
\end{figure}

The spin precession period $T_z$ as a function of excess energy
$\Delta E$ is shown in Fig.~\ref{fig:Calculation}. For a more
quantitative treatment of initial spin relaxation, we would need
to know the energy relaxation time $\tau_E$. We cannot determine
$\tau_E$ experimentally as it is shorter than the temporal
resolution of our experimental setup. Furthermore, an estimate is
hindered by the fact that $\tau_E$ depends not only on the wave
vector $k_\|$ but also on other parameters such as the number of
scattering centers, the carrier mobility and density. Therefore, a
quantitative comparison of Eq.\ (\ref{cosine}) with our
experimental results is hardly possible. Nonetheless, we can draw
the following qualitative conclusions from the above model.

First we discuss the regime of excess energies $\Delta E \le
50$~meV. For wide QWs with well widths between $198$ and $122$~{\AA}
we obtain relatively large values of $T_z \gtrsim 10$~ps. Assuming a
typical energy relaxation time $\tau_E \approx 200$~fs, Eq.\
(\ref{cosine}) yields a maximum decrease of $P_0$ of $\lesssim 3$\%
so that the influence of initial spin relaxation can be neglected
for these wide wells. Narrow QWs with well widths $\leq 71$~{\AA}
exhibit short precession times $T_z \lesssim 4$~ps. This is due to
the increase of $\langle k_z^2 \rangle$ with decreasing QW width,
which causes a larger effective field $|\Omega (\bm{k}_\|)|$
according to Eq.~(\ref{FieldVector}). Consistent with these results,
Eq.\ (\ref{cosine}) predicts a large decrease of $P_0$ of about
$50$~\% in narrow wells, in good qualitative agreement with the
experimental findings.

For excess energies $\Delta E > 50$~meV we obtain $T_z \approx
\mathrm{const}$ for QW widths $\leq 83$~{\AA}. Here the
$k_\|$-linear terms in Eq.\ (\ref{omega_max}) are compensated by the
$k_\|^3$ terms. This explains why the ratio between theoretical and
experimental values of $P_0$ is approximately constant as a function
of $\Delta E$. For the wide QWs, $T_z$ shows a decrease for $\Delta
E > 50$~meV. This is easily explained by the increasing contribution
of the $k_\|^3$ terms in Eq.\ (\ref{omega_max}). If $k_\|^2 \gg
\langle k_z^2\rangle$ we expect therefore a strong influence of
initial spin relaxation even for wide QWs. This, however, cannot be
explored further in the present work, as the energy range is beyond
what can be covered by our calculations.

Finally we note that we expect no influence of initial spin
relaxation on the measured $P_0$ in symmetric (110)-oriented GaAs
QWs since here the effective magnetic field is always pointing
perpendicular to the plane of the QW. \cite{win04, dohrmann04}
Therefore, the optically oriented electron spins are parallel to
the vector of the effective magnetic field so that the
Dyakonov-Perel spin relaxation is suppressed. We have measured
$P_0$ as a function of the excitation energy in a (110) GaAs
multiple QW structure containing 10 wedge shaped QWs. For a well
width of 47~{\AA} the confinement energy in the (110)-oriented
GaAs/Al$_{0.4}$Ga$_{0.6}$As QW is similar to the confinement
energy of the 63~{\AA} wide (100)-oriented GaAs/AlAs QW. While in
the latter QW the measured degree of polarization above the LH1:E1
exciton is rather small (Fig.~\ref{fig:AllQWs}), we have obtained
values of $P_0$ for the (110)-oriented QW which are comparable in
magnitude to the calculated spin polarization at these excitation
energies. This corroborates our conclusion that the measured
polarization is reduced because of initial spin relaxation.

\subsection{\label{sec:Density}Dependence of optical orientation on
excitation power}

\begin{figure}[t]
\includegraphics[width=1.0\linewidth]{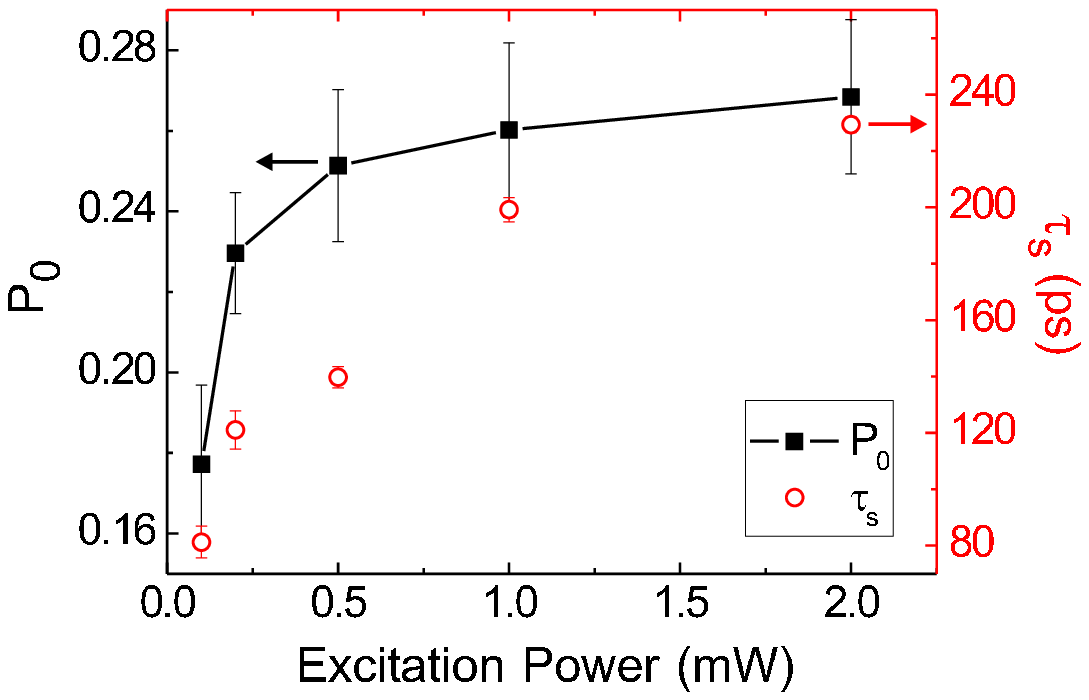}
\caption{\label{fig:Density} Initial spin
polarization $P_0$ (black filled squares) and spin lifetime
$\tau_s$ (red open circles) as a function of excitation power for
the $122$~{\AA} wide QW at an excess energy of $176$~meV. The
black line is a guide to the eye.}
\end{figure}

Figure~\ref{fig:Density} shows the measured degree of electron spin
polarization $P_0$ as a function of the excitation power for the
$122$~{\AA} wide QW at an excess energy $\Delta E = 176$~meV. We
observe a significant increase of $P_0$ for larger excitation
powers. In the following we explain this increase by more efficient
motional narrowing during initial spin relaxation.

In Sec.~\ref{subs:InitialRelaxation} we obtained a qualitative
estimate for the initial spin relaxation by evaluating the average
spin precession period $T_z$ of the optically excited electron
states prior to the first scattering event. In a more realistic
picture, we must take into account multiple scattering events, too.
Each time an electron is scattered from a state with in-plane wave
vector $\bm{k}_\|$ to a state $\bm{k}_\|'$, it is exposed to a
differently oriented effective magnetic field $\bm{\Omega}
(\bm{k}_\|')$. Frequent momentum scattering events thus reduce the
spin relaxation, which is known as motional narrowing.
\cite{slichter63, dyakonov72} There are inelastic scattering events
such as electron-phonon scattering, as well as elastic momentum
scattering events which include, e.g., electron-impurity scattering
and electron-electron scattering. While the former processes are
approximately independent of the density of excited electrons,
electron-electron scattering becomes more efficient with increasing
electron density. For low excitation powers, momentum scattering is
less efficient so that the initial spin relaxation is hardly reduced
by motional narrowing. For the parameters of Fig.~\ref{fig:Density},
we have a very short precession period $T_z \approx 2.5$~ps, see
Fig.~\ref{fig:Calculation}. The measured polarization $P_0$ is
therefore very low due to effective initial spin relaxation. When
the excitation power is increased, electron-electron scattering and
motional narrowing become more efficient. Therefore, the initial
spin relaxation is reduced and the measured spin polarization $P_0$
increases with excitation power. All data shown in
Fig.~\ref{fig:AllQWs} was obtained with an excitation power of
$1$~mW where initial spin relaxation was partly suppressed by
motional narrowing.
Of course, electron-electron scattering and motional narrowing
affect not only the initial spin relaxation but also the spin
relaxation at later times, as described by $\tau_s$ in Eq.\
(\ref{eq:polfit}). Consistent with the above arguments, we obtain
spin relaxation times $\tau_s$ which increase with excitation
power, see the open circles in Fig.~\ref{fig:Density}.
Finally we note that for the low to moderate excitation powers
considered here phase space filling of the exciton states is not
important.

\section{\label{sec:theory}Theoretical Analysis}

\subsection{Theoretical model}

Our theory for the excitonic absorption follows
Ref.~\onlinecite{Winkler95}. The main idea is to expand the exciton
wave functions in terms of electron and hole states. The exciton
Schr\"odinger equation is then solved in momentum space by means of
a modified quadrature method. Finally we calculate the
energy-dependent absorption coefficient using Fermi's Golden Rule.

For both the electron and hole states we use an $8 \times 8$ Kane
multiband Hamiltonian \cite{Trebin79} containing the lowest
conduction band $\Gamma_6^c$, the topmost valence $\Gamma_8^v$ and
the split-off valence band $\Gamma_7^v$. In the axial approximation,
\cite{win03} the single-particle states become
\begin{equation}
\label{einteilchenwf_axial}
\psi_{n \bm{k}} (\bm{r}) = \frac{1}{2 \pi}
\sum_{j} e^{i \bm{k} \cdot \bm{\rho}}
\; e^{- i M_j \varphi} \: \xi^j_{n k} (z) \, u_j (\bm{r}) \; ,
\end{equation}
where $\bm{r} = (\bm{\rho}, z)$ is the position vector and $n$ is
the subband index. In this section, $\bm{k} = (k \cos\varphi,
k\sin\varphi)$ is the in-plane wave vector, i.e., we omit the index
$\|$. The quantum number $M_j$ is the $z$ component of the angular
momentum of the $j$\/th spinor component $\xi_{n\bm{k}}^j (z)$,
i.e., in the $8 \times 8$ model used here, $M_j$ generalizes the
quantum number $S_z$ used in the preceding sections of this paper.
Finally, $u_j (\bm{r})$ are bulk band edge Bloch functions. It is
important to note that, due to the sum over $j$, the states
(\ref{einteilchenwf_axial}) are not eigenstates of angular momentum.
Only for $\bm{k}=0$ the hole states are pure HH or LH states. We
thus label hole subbands as HH or LH-like according to the dominant
spinor components at $\bm{k}=0$. Due to HH-LH mixing, we cannot
distinguish between these subbands at large wave vectors $k$.

In the following, we consider only the optically active exciton
states with center-of-mass momentum zero. Accordingly, the exciton
states depend only on the relative coordinate $\bm{\rho} =
\bm{\rho}_e - \bm{\rho}_h$, where the index $e$ ($h$) refers to
electron (hole) states. In the axial approximation, the exciton
states can be classified by $l$, the $z$ component of the total
angular momentum. The exciton states then read
\begin{widetext}
\begin{equation}
\label{exziton_wellenf}
\Psi_{l\alpha} (\bm{\rho}, z_e, z_h)
= \frac{1}{(2\pi)^{3/2}}
\sum_{n_e, \, n_h} \sum_{j_e, \, j_h}
\int d^2 k \:
\phi_{l\alpha\, k}^{n_e n_h}
\: e^{i \bm{k} \cdot \bm{\rho}}
\: e^{i (l - M_{j_e} + M_{j_h}) \varphi} \:
\xi^{j_e}_{n_e k} (z_e) \; \xi^{j_h \,\ast}_{n_h k} (z_h)
\; u_{j_e} (\bm{r}) \, u_{j_h}^\ast (\bm{r}) \,,
\end{equation}
where $\phi_{l\alpha\, k}^{n_e n_h}$ are the expansion
coefficients. The index $\alpha$ labels exciton states with the same
value of $l$. Unlike the exciton states in simplified theories (see,
e.g., Ref.~\onlinecite{Maialle93}), the exciton states
(\ref{exziton_wellenf}) cannot be written as a direct product of
electron and hole states with well-defined quantum numbers of
angular momentum. In Eq.\ (\ref{exziton_wellenf}) only $l$ and
$\alpha$ are good quantum numbers.

Using Fermi's Golden Rule, the oscillator strength of the excitons
per unit area is given by
\begin{subequations}
\label{eq:select}
\begin{equation}
\label{coulomboszi}
f_{l\alpha}^{\,\hat{\bm{e}}} =
\frac{1}{\pi \, m_0 \, E_{l\alpha}} \;
\bigg|
\sum_{n_e, \, n_h} \sum_{j_e, \, j_h}
\mathcal{P}_{l\alpha\, n_e n_h}^{j_e j_h}
\bigg|^2 \; ,
\end{equation}
where $E_{l\alpha}$ is the energy of the exciton $(l, \alpha)$, and the
components of the dipole matrix elements are
\begin{equation}
\label{eq:dipolmat}
\mathcal{P}_{l\alpha\, n_e n_h}^{j_e j_h} =
\delta_{l - M_{j_e} + M_{j_h}, \, 0} \;
  \int d k \, k \;
\phi_{l\alpha\, k}^{n_e n_h} \:
\int d z \;
\xi^{j_h \,\ast}_{n_h k} (z) \: \xi^{j_e}_{n_e k} (z) \:
\langle u_{j_h} | \,\hat{\bm{e}}\cdot \bm{p}\, | u_{j_e} \rangle \,.
\end{equation}
\end{subequations}
\end{widetext}
Here, $\bm{p}$ is the momentum operator and $\hat{\bm{e}}$ denotes
the polarization vector of the incident light. We have $\hat{\bm{e}}
= (1/\sqrt{2})(1,\pm i,0)$ for $\sigma^\pm$ polarized light. The
matrix elements $\langle u_{j_h}| \,\bm{p}\, | u_{j_e} \rangle$ are
the same as those momentum matrix elements in the $8\times 8$ Kane
Hamiltonian which are responsible for the off-diagonal $\bm{k} \cdot
\bm{p}$ coupling between conduction and valence bands. In our
theoretical model, Eq.\ (\ref{eq:select}) replaces the selection
rules depicted in Fig.~\ref{fig:selectrule}. The Kronecker $\delta$
in Eq.\ (\ref{eq:dipolmat}) is reminescent of the simple selection
rules. For circularly polarized light (polarization $\sigma^\pm$)
only excitons with $l = \pm 1$ are optically active.

The absorption spectrum is given by
\begin{equation}
\label{absorbkoeffexit}
\alpha_{\hat{\bm{e}}} \,(\omega) =
\alpha_0 \: \sum_{l,\alpha} f_{l\alpha}^{\,\hat{\bm{e}}} \:
\delta (\hbar\omega - E_{l\alpha}) \,,
\end{equation}
where $\alpha_0 = \hbar e^2 \pi / (2 m_0 \varepsilon_0 c n)$ with
$n$ the index of refraction and $\hbar\omega$ is the excitation
energy. In the numerical calculations we replace the delta functions
by a phenomenological Lorentzian broadening.

The electron spin orientation induced by the optical creation of an
exciton $(l,\alpha)$ is the expectation value of the electron spin
operator $\hat{S}_z^e$
\begin{subequations}
\label{exit_drehimpuls}
\begin{eqnarray}
M_{l\alpha}^e & = & \langle \hat{S}_z^e \rangle_{l\alpha} \\[1ex]
& = & \sum_{j_e} M_{j_e} \! \sum_{n_e, \, n_h}
\int d k \, k \;
\big| \phi_{l\alpha\, k}^{n_e n_h} \big|^2
\int d z_e \;
\big| \xi^{j_e}_{n_e k} (z_e) \big|^2 \, .
\nonumber\\[-2ex]
\end{eqnarray}
\end{subequations}
The number of optically excited excitons $(l,\alpha)$ is
proportional to the oscillator strength
$f_{l\alpha}^{\,\hat{\bm{e}}}$. Accordingly, the spin polarization
$S_e (\omega)$ of the electron systems is given by
\begin{equation}
  \label{eq:spinpoltheo}
  S_e (\omega) =
  \frac{\alpha_0}{\alpha_{\hat{\bm{e}}} \,(\omega)}
  \sum_{l,\alpha} M_{l\alpha}^e \, f_{l\alpha}^{\,\hat{\bm{e}}} \:
  \delta (\hbar\omega - E_{l\alpha}) \,.
\end{equation}
It is the quantity $S_e (\omega)$ which we compare with the measured
spin polarization $P_0$.

\subsection{Discussion}

In Sect.~\ref{sec:expresults} we demonstrated the good agreement
between the measured data and the calculated spin polarization. In
this section we will show that a detailed understanding of these
results can be achieved based on a careful examination of the
calculated spectra.

\begin{figure}[tbp]
  \includegraphics[width=0.8\linewidth]{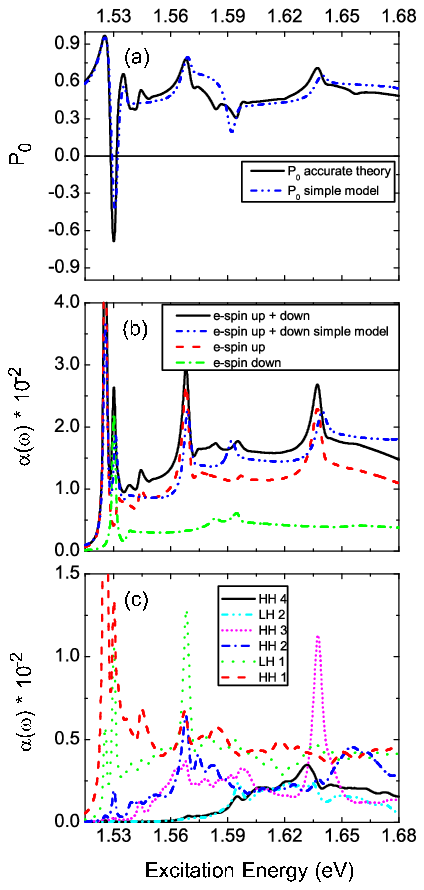}
  \caption{\label{fig:Roland} Calculated initial
  degree of polarization $P_0$ and absorption coefficient
  $\alpha(\omega)$ for circularly polarized light as
  a function of excitation energy for the
  198~{\AA} wide GaAs/AlAs QW.  The black lines in (a) and
  (b) show $P_0$ and $\alpha(\omega)$ for the full theory, whereas
  the dashed-double-dotted lines show for comparison the results for
  a simplified model that neglects valence band mixing and HH-LH
  coupling. The red dashed line (green dash-dotted line) in (b)
  shows
  the partial absorption spectra for spin up (spin down) electrons.
  Panel (c) displays the hole subband contributions
  (\ref{oszi_subband}) to $\alpha(\omega)$ for the full theory.}
\end{figure}

As an example, we show in Fig.~\ref{fig:Roland}(a) the calculated
electron spin polarization $S_e (\omega)$ and in
Fig.~\ref{fig:Roland}(b) the absorption coefficient $\alpha
(\omega)$ for the $198$~{\AA} wide QW, see also Fig.~\ref{fig:198}.
Frequently the interpretation of excitonic spectra is based on the
simple and intuitive idea that the excitons giving rise to the peaks
in the spectra can be attributed to pairs of individual electron and
hole subbands. However, such a scheme must be used with caution
because the spectra are often strongly affected by valence band
mixing and Coulomb coupling between subbands. \cite{Winkler95} To
illustrate the importance of these effects, the dashed-double-dotted
lines in Figs.~\ref{fig:Roland}(a) and~\ref{fig:Roland}(b) show the
results of a simplified calculation that neglects these couplings.
Both the absorption coefficient and the initial spin polarization
differ remarkably in these models. In particular, we find that the
oscillator strength of the HH3:E1(1s) exciton is by a factor $\sim
50$ smaller if these couplings are neglected so that the peak cannot
be resolved on the scale of Fig.~\ref{fig:Roland}. Furthermore, the
peaks labeled HH2:E2(1s) and HH3:E3(1s) are shifted to higher
energies.

\begin{figure}[tbp]
\includegraphics[width=0.8\linewidth]{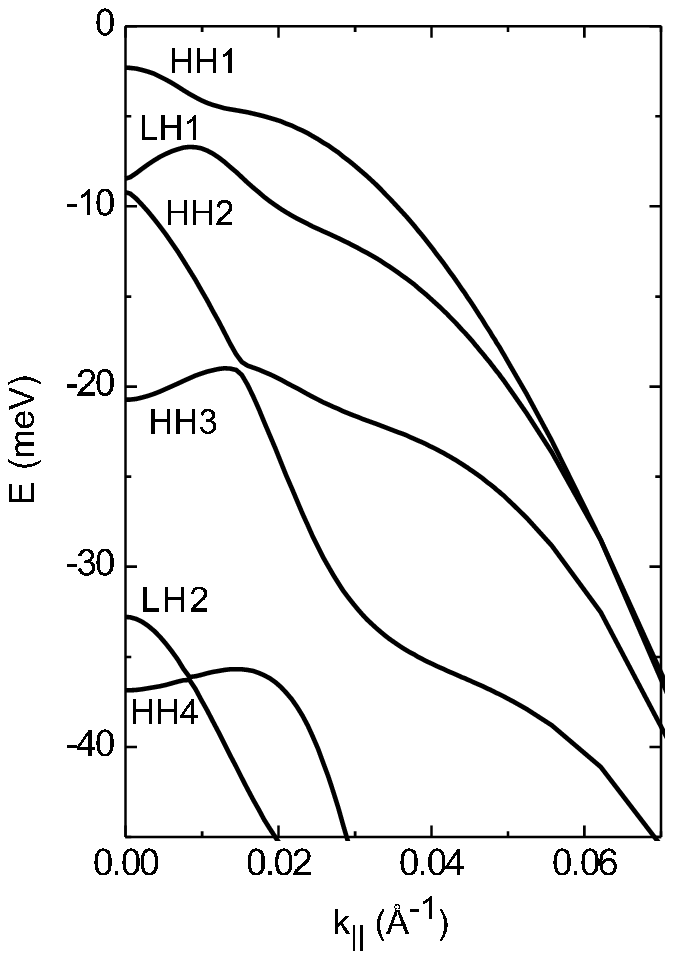}
\caption{\label{fig:subband198}  Hole subband dispersion curves
calculated for the 198~{\AA} wide QW.}
\end{figure}

In order to quantify the valence band mixing, we can evaluate the
contribution of different hole subbands to the oscillator strengths
$f_{l\alpha}^{\,\hat{\bm{e}}}$. We define the partial oscillator
strengths
\begin{subequations}
\label{oszi_subband}
\begin{equation}
f_{l\alpha,n_h}^{\,\hat{\bm{e}}} =
\frac{1}{\mathcal{N}} \,
\bigg|
\sum_{n_e} \sum_{j_e, \, j_h}
\mathcal{P}_{l\alpha\, n_e n_h}^{j_e j_h}
\bigg|^2 \; ,
\end{equation}
where the normalization $\mathcal{N}$ is chosen such that we have
\begin{equation}
\label{oszi_subband_norm}
\sum_{n_h} f_{l\alpha,n_h}^{\,\hat{\bm{e}}}
= f_{l\alpha}^{\,\hat{\bm{e}}} \; .
\end{equation}
\end{subequations}
Similar to Eq.\ (\ref{absorbkoeffexit}) we then calculate partial
spectra showing the contributions of each hole subband to the
absorption coefficient, see Fig.~\ref{fig:Roland}(c). The
complicated curves clearly illustrate that the labeling in terms of
subbands is very problematic. For example, the contribution of the
HH1 subband to the oscillator strength of the LH1:E1(1s) exciton is
larger than the contribution of the LH1 subband. For comparison, we
show in Fig.~\ref{fig:subband198} the hole subband dispersion curves
of the 198~{\AA} wide QW.

We suggest here a different approach for decomposing the spectra
that yields a much clearer physical picture. We can identify whether
the oscillator strength of an exciton is predominantly from the
dipole matrix element (\ref{eq:dipolmat}) between a hole and a
spin-up or a spin-down electron state by evaluating the partial
oscillator strengths
\begin{equation}
\label{oszi_spinor}
f_{l\alpha,j_e}^{\,\hat{\bm{e}}} =
\frac{1}{\mathcal{N}} \,
\bigg|
\sum_{n_e, \, n_h} \sum_{j_h}
\mathcal{P}_{l\alpha\, n_e n_h}^{j_e j_h}
\bigg|^2 \; ,
\end{equation}
where the normalization $\mathcal{N}$ is choosen analogously to Eq.\
(\ref{oszi_subband_norm}). We then calculate partial spectra for the
spin-up and spin-down spinor components $j_e$, see the dashed and
dash-dotted lines in Fig.~\ref{fig:Roland}(b). For the different QWs
investigated in this work, we show the partial oscillator strengths
(\ref{oszi_spinor}) in the bottom panels of Figs.~\ref{fig:198}
and~\ref{fig:AllQWs}.

For the eight-component spinors (\ref{einteilchenwf_axial}) we
obtain eight partial oscillator strengths (\ref{oszi_spinor}).
However, for the electron states, the contributions of the valence
band spinor components are very small so that they could not be
resolved using the scale of Fig.~\ref{fig:Roland}. (Yet these spinor
components are very important for the correct absolute values of the
exciton energies. \cite{Winkler95}) The partial oscillator strengths
$f_{l\alpha,j_e}^{\,\hat{\bm{e}}}$ for the spin-up and spin-down
components of the electron states are essentially equivalent to the
corresponding partial oscillator strengths
$f_{l\alpha,j_h}^{\,\hat{\bm{e}}}$ of the HH and LH components of
the hole states, see Fig.~\ref{fig:selectrule}. These partial
oscillator strengths would be strictly equal in a $6 \times 6$ model
that neglects the split-off valence band $\Gamma_7^v$.

Unlike for the partial oscillator strengths (\ref{oszi_subband}), we
get from Eq.\ (\ref{oszi_spinor}) a clear and simple decomposition
of the spectra. In particular, a comparison between the partial
spectra in Fig.~\ref{fig:Roland}(b) and the electron spin
polarization in Fig.~\ref{fig:Roland}(a) shows that each resonance
can be labeled as an excitation of either spin-up or spin-down
electrons, consistent with Fig.~\ref{fig:selectrule}. In spite of
the strong admixture of different hole subbands visible in
Fig.~\ref{fig:Roland}(c) it is either the electron spin-up \emph{or}
the spin-down component (i.e., the HH \emph{or} the LH component) of
an exciton that is optically active. The reason why we get much
clearer results from Eq.\ (\ref{oszi_spinor}) than from Eq.\
(\ref{oszi_subband}) lies in the fact that the labeling of hole
subbands as HH- or LH-like (see Fig.~\ref{fig:subband198}) is not
rigorously justified, but it reflects merely the dominant spinor
component around $k=0$. For larger in-plane wave vectors, the
subbands are strongly affected by HH-LH mixing. Yet the excitons
(\ref{exziton_wellenf}) ``try to avoid the HH-LH mixing by selecting
the spinor components as a function of $k$ from different hole
subbands.'' This is also the reason why we can label most of the
excitonic resonances by pairs of electron and hole subbands
(Fig.~\ref{fig:198}). This scheme refers to the pairs of electron
and hole subbands that contribute the largest around $k=0$. At
larger wave vectors $k$ in the expansion (\ref{exziton_wellenf}),
the exciton states contain large contributions from other subbands,
too.

\begin{figure}[tbp]
\includegraphics[width=0.8\linewidth]{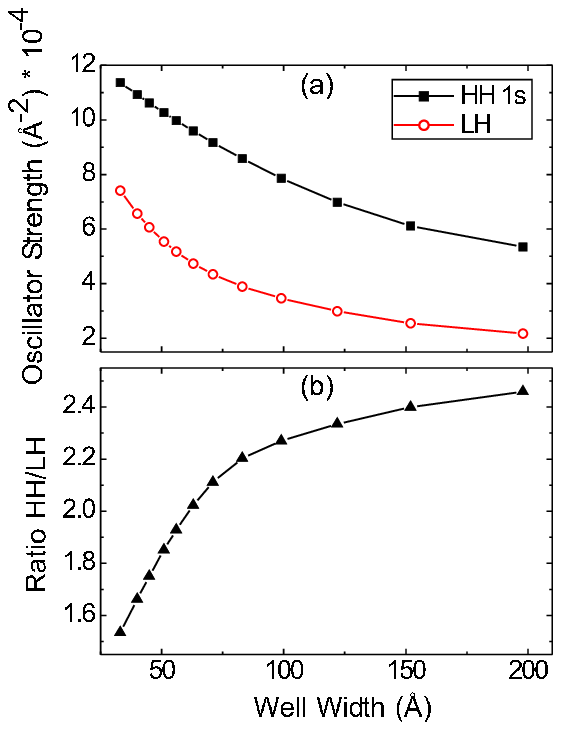}
\caption{\label{fig:oscillatorstrength} Calculated
oscillator strength (a) for the HH and LH transitions as a
function of the QW width. The ratio between the HH and LH
oscillator strength is shown in (b).}
\end{figure}

In spite of the fact that we can label the excitonic resonances by
pairs of electron and hole subbands, the oscillator strengths of the
individual resonances in Fig.~\ref{fig:Roland}(b) are very different
from those in Fig.~\ref{fig:selectrule}. To illustrate this point,
Fig.~\ref{fig:oscillatorstrength}(a) shows the oscillator strength
of the HH1:E1(1s) and the LH1:E1(1s) exciton as a function of well
width while Fig.~\ref{fig:oscillatorstrength}(b) shows the ratio
between these quantities. Only in the limit of very wide QWs we
approach the bulk value 3. The ratio decreases with decreasing well
width due to HH-LH coupling. For energies larger than the HH1:E1
absorption edge ($E > 1.535$~eV for the $198$~{\AA} wide QW), the
individual peaks in the spectra are Fano resonances, i.e., they are
degenerate with the continua of excitons from lower subband pairs.
The peaks on top of the continua are thus less important for the
electron spin polarization observed at these energies. The magnitude
of the electron spin polarization in this regime is always smaller
than one.

The good agreement between the theory and the experimental data has
been demonstrated in Figs.~\ref{fig:198} and~\ref{fig:AllQWs}. It
indicates that our basic assumption $P_\mathrm{opt} = P_s$ is
justified. This implies that the orientation of the electron spins
is essentially preserved when the electrons relax from the optically
excited high-energy states to a thermal equilibrium for their
momentum distribution. (Only in narrow QWs the measured optical
polarization is smaller than the calculated electron spin
polarization due to initial spin relaxation.)

At a first glance, our findings suggest that in our experiments
electrons and holes relax independently from the optically excited
state to quasi-thermal equilibrium, the reason being that in a
single-particle picture the spin of the electrons is a good
quantum number. \cite{almost} On the other hand, only excitons
with angular momentum quantum number $l=\pm 1$ can absorb or emit
photons with polarization $\sigma^\pm$. With $\sigma^+$ polarized
light we thus excite only excitons with angular momentum $l=+1$.
If the excitons preserved the angular momentum quantum number $l$
while they relax from the optically excited states to thermal
equilibrium, the measured optical polarization $P_\mathrm{opt}$
would be the same like the polarization of the exciting laser
beam, independent of the energy of the laser. This disagrees
clearly with our experimental findings. We note, however, that
each dublet of optically active excitons with $l=\pm 1$ is almost
degenerate with a dublet of optically inactive excitons with
$l=\pm 2$ or $l=0$ (Refs.~\onlinecite{Winkler95, exchange}). The
latter dublet is related to the optically active excitons by a
spin flip of the hole. The electron spin of the exciton state (but
not $l$) can thus be preserved even if the hole spin of the
exciton is flipped. Therefore, we cannot decide, based on our
experiments whether electrons and holes relax independently or
whether they relax as a Coulomb-correlated exciton state.
Recently, two groups were able to gain information on exciton
formation dynamics in GaAs quantum wells by using optical-pump
THz-probe spectroscopy and time-resolved PL on a very high quality
quantum well.\cite{kaindlNATURE03,szczytkoPRL04}

\section{\label{sec:conclusion}Conclusions}

Using time resolved photoluminescence excitation spectroscopy and a
multiband envelope function theory of excitonic absorption based on
the $8 \times 8$ Kane Hamiltonian, we have studied the energy
dependence of the initial degree of spin polarization of optically
created electrons in GaAs QWs with different well widths. Taking
into account Coulomb coupling and HH-LH coupling between subbands
was shown to be essential to obtain good agreement between theory
and experiment for a wide range of excitation energies. The
calculated results differ significantly from the experimental data
if a frequently used simplified exciton model is applied that
neglects these couplings. This work therefore provides the first
quantitative picture of the optical orientation of electron spins in
GaAs QWs.

The good agreement between the measured degree of optical
polarization $P_\mathrm{opt}$ and the calculated spin polarization
$P_s$ of the electrons indicates that our basic assumption
$P_\mathrm{opt} = P_s$ is justified. This implies that the
orientation of the electron spins is (essentially) preserved when
the electrons relax from the optically excited high-energy states to
a thermal equilibrium for their momentum distribution. In narrow QWs
the measured optical polarization is smaller than the calculated
electron spin polarization due to initial spin relaxation. However,
this process is found to be essentially independent of the energy of
the exciting photons. Initial spin relaxation is most effective for
small excitation powers. For larger excitation powers it becomes
less important because of motional narrowing.


\begin{acknowledgments}
  This work was supported support by BMBF and DFG. S.~P.\ thanks the
  Friedrich-Ebert-Stiftung for financial support.
\end{acknowledgments}


\begin{thebibliography}{34}
\bibitem[{alm()}]{almost}
\bibinfo{note}{Strictly speaking, the spin of the electrons is an almost good
  quantum number because of the effective magnetic field (\ref{FieldVector})
  which gives rise to spin relaxation.}
\bibitem[{exc()}]{exchange}
\bibinfo{note}{The dublet of optically active excitons and the dublet of
  inactive excitons are split by the electron-hole exchange interaction.}
\end{thebibliography}


\end{document}